\documentstyle[12pt]{article}
\begin{document}
\setlength{\baselineskip}{20pt}
\begin{center}
{\Large{\bf Spinning cosmic strings: a general class of solutions}}
\\[20mm]
{\large N. \" Ozdemir} \\[2mm]
ITU, Faculty of Sciences and Letters, Department of Physics,\\
34469 Maslak, Istanbul, Turkey \\
and\\
Feza G\"ursey Institute, P.K. 6  \c Cengelk\" oy,
81220 Istanbul, Turkey \\
\end{center}
\vfill
\begin{abstract}
In this work, we give a general class of solutions of the spinning
cosmic string in Einstein's theory of gravity. After treating same
problem in Einstein Cartan (EC) theory of gravity, the exact
solution satisfying both exterior and interior space-times
representing a spin fluid moving along the symmetry axis is
presented in the EC theory. The existence of closed timelike curves
in this spacetime are also examined.
\end{abstract}
\newpage
\section{Introduction}

Cosmic strings are line-like objects. Like other topological defects
\cite{vilsh}, they might have been formed during the phase
transitions in the very early universe.  The cosmic string
space-time is locally flat. Globally it has a conical structure that
may result in some non-trivial global effects.  The current
experimental results exclude the predictions of the cosmic strings
about the large scale structure and the anisotropy in the cosmic
microwave background radiation.  Their gravitational effects,
however, make it worthwhile to study cosmic strings.  These effects
may be listed as: vacuum fluctuations, gravitational lensing,
bending of light.  They may also give an explanation to the most
energetic events in the universe such as ultrahigh energy neutrinos,
cosmological gamma-ray bursts.  Furthermore, a mixture of
topological effects and inflation still gives a theory consistent
with the current cosmic microwave background data \cite{pogvach}.

A straight cosmic string can be characterized by its linear mass
density $\mu$ which depends on the formation energy scale.  For the
cosmic strings in the grand-unified theories, the linear mass
density is $\mu= 10^{22} gr/cm .$  Here we are interested in a
special class of cosmic strings, namely the spinning cosmic strings
which can be determined by their linear energy density $\mu$ and
intrinsic spin $J$. Geometrically the spinning cosmic string
space-time can be obtained by cutting a wedge from four dimensional
Minkowski space-time and gluing it back after making a boost in one
of the faces. This boost is different from the one made in the
"straight string" case. Since there is no Lorentz boost invariance
anymore, an observer moving along the $z$-axis will see a twisted
metric with the spacelike helical structure [3-6]. In the spinning
cosmic string spacetime, some physical properties are studied in the
literature like the quantization of the energy \cite{mazur},
semi-classical gravitational effects \cite{mastas}, vacuum
fluctuations \cite{lormor, lorpaosva} and closed timelike curves
\cite{bonnor}.

In the following section, we present a general class of solutions
for the field equations of the Einstein's gravitational theory in
the vacuum for the spinning string. We start our calculation using
the metric proposed by \cite{gallet}, where the strings with helical
structure in the time coordinate as well as the space coordinate
along the string is presented. It is also shown that these solutions
can be proposed by considering thick cylindrical sources which do
not have $\delta$-type singularity in the curvature and
energy-momentum tensor.

In either method, other singular and non-singular sources can also
be considered such as electric current carrying strings on which
small structures (wiggles) exist.  But here spinning fluid is
considered as the reason for the singularities in the curvature and
torsion.

In the continuum mechanical point of view, a cosmic string can be
represented by disclinations in the space-time and the helical
structure of the spinning cosmic string is associated with the
timelike and spacelike dislocations.  In \cite{let}, by introducing
torsion along the string or treating the problem in the EC (
Einstein-Cartan ) theory, some ill-defined quantities are removed,
or the missing part of the spacetime defects are recovered. Since
torsion is coupled to the theory by means of spin, it is seen that,
for the spacetime presented, the torsion tensor has the same
$\delta$-type singularities as the curvature and the energy-momentum
tensor \cite{hehl}.

In the framework of the distorted spacetimes some exact solutions of
the EC theory, which can be interpreted as spin-polarized cosmic
strings and cosmic dislocations, are given in \cite{punsol}. One may
also interpret these solutions as those that exist in the presence
of sources which have non-zero thickness such as spinning fluids. In
section 3, we suppose that a spin fluid exists in a cylinder with
radius $r_0$ moving along the z-axis and fullfilling the exact
solution conditions for the EC theory.  We consider the interior (
the spin fluid with matter and torsion distribution) and exterior (
vacuum of the spin fluid) space-times induced by the spinning fluid.
In this example the interior and the exterior space-times are
subject to the continuity and the Arkuszewski-Kopczynski-Ponomariev
(AKP) junction conditions \cite{akp}, i.e.,  continuity in the
metric components and discontinuity in the first derivatives of the
metric components.

We also study closed timelike curve (CTC) properties of the metrics
corresponding to the spinning cosmic strings.

\section{Spinning cosmic string in Einstein theory of gravity}
Spinning cosmic string spacetime line element can be given by
\begin{equation}
ds^2=-(dt+4 G J^t d\phi)^2+dr^2+ b^2 r^2 d\phi^2+(dz+4 G J^z d\phi)^2
\label{spin}
\end{equation}
where $b$ is the string parameter and  defined in terms of string
energy density $\mu$ and Newton's constant $G$ as $b=1-4G\mu$. The
constants $J^t$ and $J^z$ are spin of the string and dislocations of
the space respectively \cite{gallet, let}. In these references they
examine the physical structures of the space-times depending on the
three different values of  $j^2=(J^t)^2-(J^z)^2$:

\noindent
{\bf i)}  $j^2=0$ (or $\vert J^z\vert =\vert J^t\vert$),
\begin{equation}
ds^2=-(dt+4 G J d\phi)^2+dr^2+ b^2 r^2 d\phi^2+(dz+4 G J d\phi)^2
\end{equation}
represents a string interacting with a circularly polarized plane-fronted
gravitational wave,

\noindent
{\bf ii)}  $\vert J^z\vert=0$,
\begin{equation}
ds^2=-(dt+4 G J^t d\phi)^2+dr^2+ b^2 r^2 d\phi^2+ dz^2
\end{equation}
represents a spinning cosmic string with no cosmic dislocations
and

\noindent
{\bf iii)}  $\vert J^t\vert=0$,
\begin{equation}
ds^2=-dt^2+dr^2+ b^2 r^2 d\phi^2+ (dz+4 G J^z d\phi)^2
\end{equation}
here the spacetime represents screw dislocations; the superposition of screw  dislocation
($2 GJ^z/\pi$ is analogous to Burgers vector) and disclination.

Now we consider a general vacuum solution of the Einstein field
equations corresponding to the  spinning cosmic string lying along
the $z$-axis which has the line element
\begin{equation}
ds^2=-[dt+F(z)d\phi]^2+dr^2+ b^2 r^2 d\phi^2+[dz+H(t) d\phi]^2
\label{genspin}
\end{equation}
where $F(z)$ is spin of the string and $H(t)$ is
the analogue of the Burgers vector of dislocations.
Vacuum field equations require that $F$ and $H$ must be linear in their variables such that
$F(z)=a z+J^t$and $H(t)=a t+J^z$ with $a$, $J^t$ and $J^z$ are constants.
The space-time given by (\ref{spin}) is a special
form of (\ref{genspin}), if we take $a=0$ the previous case is
obtained \cite{gallet, let}.
Since the nature of the problem will not change, for the sake of the simplicity we
make transformations $z\rightarrow (z-J^t/a)$ and $t\rightarrow (t-J^z/a)$
and take $a=4GJ$, then we get
\begin{equation}
ds^2=-[dt+4GJ z d\phi]^2+dr^2+ b^2 r^2 d\phi^2+[dz+4GJ t d\phi]^2.
\label{genspin2}
\end{equation}
Since the space-time is locally flat
except along the $z$-axis, to examine the structure of the
singularity we also need to make the transformation $r\rightarrow r^b/b$ and
write (\ref{genspin2}) in Cartesian coordinates
\begin{equation}
ds^2=-[dt+4\,G\,J z (X dy-Y dx)]^2+r^{-8 G\mu}(dx^2+ dy^2)+[dz+4\,G\,J t(X dy-Y dx)]^2,
\label{spin1}
\end{equation}
where $X={x/ r^2}$, $Y={y/ r^2}$ and $r=\sqrt{x^2+y^2}$.
In this spacetime one forms are defined as
\begin{eqnarray}
\theta^0&=& dx^0+4\,G\,J\,x^3 (X dx^2-Y dx^1)\nonumber\\
\theta^3&=& dx^3+4\,G\,J\,x^0 (X dx^2-Y dx^1)\nonumber\\
\theta^1&=& r^{-4\,G\mu}dx^1\nonumber\\
\theta^2&=& r^{-4\,G\mu}dx^2
\end{eqnarray}
$(0,3,1,2)$ represents $(t,z,x,y)$ and metric is defined as
$\,\eta_{ij}=diag(-1,1,1,1)$
and we use the definition
$\partial X/\partial x^1+\partial Y/\partial x^2=2\pi\delta(x^1)\delta(x^2)$,
same notation given in \cite{gallet, let}.

During the calculations of Einstein's tensors, in addition to
$\delta$-type distributions we obtain $(\delta(x^1)\delta(x^2))^2$
type singularities. We can approach this problem in two ways: First,
$[\delta(x^1)\delta(x^2)]^2$ term is taken to be zero or lower order
than $\delta(x^1)\delta(x^2)$ as examined in \cite{gallet,let} and
second, the metric (\ref{genspin}) represents the exterior spacetime
of the thick non-zero source and the interior solution and matching
conditions should be satisfied \cite{bekenstein}.

In the first approach the non-zero components of the Einstein  tensor are
\begin{eqnarray}
G_{00}&=&-G_{33}=8 \pi G\mu \, r^{8 G\mu}\delta(x^1)\delta(x^2) \nonumber\\[.2cm]
G_{01}&=&-G_{31}=4\pi G J x^3 \, r^{12 G\mu}\partial[\delta(x^1)\delta(x^2)]/\partial x^2\nonumber\\[.2cm]
G_{02}&=&-G_{32}=-4\pi G J x^0 \, r^{12
G\mu}\partial[(\delta(x^1)\delta(x^2)]/\partial x^1 \label{entor0}
\end{eqnarray}
%
and the scalar curvature is
\begin{equation}
R=16\pi G\mu\, r^{8 G\mu}\delta(x^1)\delta(x^2).
\end{equation}
We see that the spacetime given by (\ref{spin1}) represents a cosmic
string since tension of the string is equal to its linear energy
density i.e., $G^0_0=G^3_3$ and it is valid both for exterior
spacetime and along the singular line source \cite{his, vil2}.

In the second approach, we consider  thick string, a non-singular
source,  as an interior solution which is cylindrically symmetric.
The exterior metric is (\ref{spin1})  and satisfies matching
conditions; continuity in the metric components and junction
condition in the exterior curvatures of exterior and interior
spacetimes on the cylinder with radius $r_0$ \cite{israel}. Now we
introduce interior metric as
\begin{equation}
ds^2=-(dt+A\,d\phi)^2+dr^2+W^2d\phi^2+(dz+B\,d\phi)^2
\label{int1}
\end{equation}
where $A=c z+q(r)$ is function of $z$ and $r$, $B=c t+k(r)$ is function of
$t$ and $r$ and $W$ is function of $r$ only.

The interior spacetime can be defined by the anisotropic fluid
moving along the string ($z$-axis) and the energy momentum tensor of
an anisotropic fluid in the Riemannian (torsionless) space is given
by
\begin{equation}
t_{ij}=(\epsilon+p)u_iu_j-p\,\eta_{ij}+(P-p)v_iv_j
\label{enn0}
\end{equation}
Here, $\bf u$ is the fluid's velocity, $p$ is the isotropic pressure
in the $r$ and $\phi$ directions and $P$ is the pressure in the
$z$-direction. $\bf v$ is a spacelike vector orthogonal to ${\bf u}$
and $\epsilon$ is the  linear density of the string. Velocities obey
the normalization conditions $u^iu_i=-1$, $v^iv_i=1$ and the
orthogonality condition $u^iv_i=0$. Then, corresponding equations
become
\begin{eqnarray}
G_{00}&=&-({W^{\prime\prime}\over W}-{q^{\prime 2}\over 2W^2})=(\epsilon+p)u_0u_0+p+(P-p)v_0v_0\nonumber\\[.2cm]
G_{33}&=&({W^{\prime\prime}\over W}+{k^{\prime 2}\over 2W^2})=(\epsilon+p)u_3u_3-p+(P-p)v_3v_3\nonumber\\[.2cm]
G_{11}&=&({q^{\prime 2}\over 4W^2}-{k^{\prime 2}\over 4W^2})=-p+(P-p)v_1v_1\nonumber\\[.2cm]
G_{22}&=&({q^{\prime 2}\over 4W^2}-{k^{\prime 2}\over 4W^2})=-p+(P-p)v_2v_2\nonumber\\[.2cm]
G_{01}&=&-{k^\prime c\over 2 W^2}=(P-p)v_0v_1\nonumber\\[.2cm]
G_{02}&=&({q^\prime \over 2 W})_{,r}=(P-p)v_0v_2\nonumber\\[.2cm]
G_{31}&=&{q^\prime c\over 2 W^2}=(P-p)v_3v_1\nonumber\\[.2cm]
G_{32}&=&-({k^\prime \over 2 W})_{,r}=(P-p)v_3v_2\nonumber\\[.2cm]
G_{03}&=&-{q^\prime k^\prime \over 2
W^2}=(\epsilon+p)u_0u_3+(P-p)v_0v_3
 \label{entor0}
\end{eqnarray}
If we take $q(r)=constant$ and $k(r)=constant$ the interior solution corresponds
to spinning cosmic string where $G^0_0=G^3_3$. In general, it
represents spinning string with the Einstein's tensors given by (\ref{entor0}).
And matching conditions on the cylinder with radius $r_0$ ($t=t_0$ and $z=z_0$) are
\begin{eqnarray}
g_{\mu\nu}\vert_+&=&g_{\mu\nu}\vert_-\nonumber\\[.2cm]
K^{+}_{\mu\nu}&=&K^{-}_{\mu\nu}
\label{subcon}
\end{eqnarray}
where $_+ (_-)$ represents exterior (interior) space-time and
$K^\pm_{\mu\nu}$ represents the corresponding exterior derivative,
$K^\pm_{ij}=e^\alpha_i e^\beta_j\,n^{\pm}_{\alpha;\beta}$.
$e^\alpha_i$ is the orthonormal triad lying in the junction surface
and $n^\pm_\alpha$ is the unit normal vector outward (inward) for
the interior (exterior) space-times. These conditions give the
following set of equations
\begin{eqnarray}
-az_0^2+a^2t_0^2+b^2 r_0^2&=&[c z_0+q(r_0)]^2+[c t_0+k(r_0)]^2+W^2(r_0)\nonumber\\[.2cm]
az_0&=&cz_0+q(r_0)\nonumber\\[.2cm]
at_0&=&ct_0+k(r_0)\nonumber\\[.2cm]
b^2r_0&=&qq^\prime+kk^\prime+WW^\prime\nonumber\\[.2cm]
0&=&k^\prime\nonumber\\[.2cm]
0&=&q^\prime
\label{jun0}
\end{eqnarray}
from last two equations it is seen that $q$ and $k$ must be
functions of $(r-r_0)$ or equal to constants. We can make
assumptions $W(r)=\sin r$, $q=\sin (r+\pi/2)$ and $k= r+const.$ such
that our solution will represent fluid moving in a cylinder with
radius $r_0$ which is similar to the straight string solution but a
little complicated. Corresponding $r$ dependent velocities are as
follows
\begin{eqnarray}
u_0&=&{1\over \sqrt{\epsilon+p}}\left(-p +{q^{\prime 2}\over 2
W^2}-{c^2 k^{\prime 2}\over 2 W^2(2 p W^2-k^{\prime 2}+q^{\prime
2})}\right)^{1/2}\nonumber\\[.2cm]
u_3&=&{1\over \sqrt{\epsilon+p}}\left(p +{k^{\prime 2}\over 2
W^2}-{c^2 q^{\prime 2}\over 2 W^2(2 p W^2-k^{\prime 2}+q^{\prime
2})}\right)^{1/2}\nonumber\\[.2cm]
v_0&=&-{c k^\prime\over W\sqrt 2(P-p)}\left(2 pW^2-k^{\prime
2}+q^{\prime 2}\right)^{1/2}\nonumber\\[.2cm]
v_3&=&{c q^\prime\over W\sqrt 2(P-p)}\left(2 pW^2-k^{\prime
2}+q^{\prime 2}\right)^{1/2}\nonumber\\[.2cm]
v_1&=&{1\over\sqrt{P-p}}\left(p-{k^{\prime 2}\over 2 W^2}+{
q^{\prime 2}\over 2 W^2}\right)^{1/2}\nonumber\\[.2cm]
v_2&=&{1\over\sqrt{P-p}}\left(p-{k^{\prime 2}\over 2 W^2}+{
q^{\prime 2}\over 2 W^2}\right)^{1/2}.
\end{eqnarray}
with the constraints coming from last two equations of
(\ref{entor0}) are
\begin{eqnarray}
{c q^\prime-k^\prime W^\prime+W k^{\prime\prime}\over 2 W^2}&=&0\nonumber\\[.2cm]
{c k^\prime+q^\prime W^\prime+W q^{\prime\prime}\over 2 W^2}&=&0,
\end{eqnarray}
and from the orthogonality of $\bf{u.v}=0$ we get
\begin{equation}
k^\prime\sqrt{-2 p W^2+q^{\prime 2}+{c^2 k^{\prime 2}\over -2
pW^2+k^{\prime 2}-q^{\prime 2}}}+q^\prime\sqrt{2 p W^2+k^{\prime
2}+{c^2 q^{\prime 2}\over -2 pW^2+k^{\prime 2}-q^{\prime 2}}}=0.
\nonumber\\[.2cm]
\end{equation}
Pressures can be obtained from the normalization conditions of
$\bf{u.u}=-1$ and $\bf{v.v}=1$
\begin{eqnarray}
{8 p^2 W^4+2 p W^2(-k^{\prime 2}+q^{\prime 2})+(k^{\prime
2}-q^{\prime 2})(c^2-k^{\prime 2}+q^{\prime 2})\over 2(\epsilon+p)
W^2 (2 p W^2-k^{\prime 2}+q^{\prime 2})}&=&-1\nonumber\\[.2cm]
{-8 p^2 W^4+8 p W^2(k^{\prime 2}-q^{\prime 2})+(k^{\prime
2}-q^{\prime 2})(c^2-2 k^{\prime 2}+2 q^{\prime 2})\over 2(p-P) W^2
(2 p W^2-k^{\prime 2}+q^{\prime 2})}&=&1
\end{eqnarray}
and they should be equal to zero on the boundary surface $p=P=0$ at
$r=r_0$. If we choose $W(r)=b r$ or $W(r)=b \sin(r/b)$ as in the
literature, we meet non-simple  choice of $q(r)$ and $k(r)$
therefore non-simple pressures but, one can always find solutions
desired functions for $W$, $q$, $k$ and/or pressures.

\section{Spinning cosmic string in Einstein-Cartan theory}
We can treat the same problem; the general form of the spinning
cosmic string in the framework of EC-theory of gravity in which  the
torsion is different from zero and coupled to the theory through the
spin momentum density of the matter. The  field equations to be
satisfied in EC theory  are
\begin{eqnarray}
R_{ij}-{1\over 2}g_{ij} R=\kappa t_{ij}\label{tor1}\\
T^i_{jk}-\delta^i_j T^l_{lk}-\delta^i_k T^l_{jl}=\kappa s^i_{jk}
\label{tor2}
\end{eqnarray}
where $\kappa$ is gravitational constant and we take $-8\pi G$.
Here, $t_{ij}$ energy-momentum, $T^i_{jk}$ torsion  and $s^i_{jk}$
spin tensors of the matter. From, (\ref{tor2}) it is seen that the
torsion is coupled to the system by  means of spin distribution of
the matter. Here, $s^i_{jk}=u^i S_{jk}$, $u^i$ is 4-velocity  and
$S_{jk}$  intrinsic-spin angular momentum of the matter which are
subject to the Frenkel condition $u^i S_{ji}=0$.

In this section we try to find a spacetime of EC theory which
satisfies (\ref{tor1}),(\ref{tor2}). Let us suppose that the
spacetime (\ref{genspin}) represents a spin fluid flowing along the
$z$-axis with velocity $u=(u^0,u^3,0,0)$ then spin fluid moving
along the $z$-axis generates torsion distribution as
$T^0=T^0_{12}\,\theta^1\,\wedge\theta^2$,
$\,\,T^3=T^3_{12}\,\theta^1\wedge\theta^2$. Now we consider two
different torsions

\noindent {\bf i)}
\begin{eqnarray}
T^0_{12}&=&8\pi\,G\,J\,r^{8 G\mu}(x^3-1)\delta(x^1)\delta(x^2)\nonumber\\
T^3_{12}&=&8\pi\,G\,J\,r^{8 G\mu}(x^0-1)\delta(x^1)\delta(x^2)
\label{tt1}
\end{eqnarray}
which has the same functional form with (\ref{entor0})
and

\noindent {\bf ii)}
\begin{eqnarray}
T^0_{12}&=&8\pi\,G\,J\,r^{8 G\mu}\,x^3\,\delta(x^1)\delta(x^2)\nonumber\\
T^3_{12}&=&8\pi\,G\,J\,r^{8 G\mu}\,x^0\,\delta(x^1)\delta(x^2)
\label{tt2}
\end{eqnarray}
which gives the same energy-momentum with the straight cosmic string with non-zero torsion,
\begin{equation}
G_{00}=-G_{33}=8 \pi G\mu\, r^{8 G\mu}\delta(x^1)\delta(x^2).
\label{entor2}
\end{equation}

The non-zero components of the spin tensor which are the solutions of (\ref{tor2})
are $s^0_{12}=-T^0_{12}/8\pi G$ and $s^3_{12}=-T^3_{12}/8\pi G$,
and non-singular spin angular momentum
is $S_{12}=s^0_{12}/u^0=s^3_{12}/u^3$.

In Einstein theory of gravity, the space-time is torsionless and the
curvature and non-zero components of the energy momentum tensor have
$\delta$-type singularities. In EC theory in addition to these
singularities we see that the torsion tensor has the same type of
singularity.

Torsion distributions given by  (\ref{tt1}) and (\ref{tt2}),
and corresponding energy-momentum densities given by (\ref{entor0}) and (\ref{entor2})
defines a spinning cosmic string both exterior and along the singular source with torsion.

Our aim is to find the exact solution of the field equations both
satisfied by vacuum solution (\ref{genspin}) as exterior, and the
spin fluid as interior. Now we introduce cylindrically symmetric
interior metric as
\begin{equation}
ds^2=-[dt+A(t,z,r)d\phi]^2+dr^2+ W^2(r) d\phi^2+[dz+B(t,z,r) d\phi]^2
\end{equation}
with $A=f(t,z)+q(r)$, $B=h(t,z)+k(r)$ and the spin fluid moving
along the z-axis produces the torsion
\begin{eqnarray}
T^0_{12}&=&A^\prime/W=q^\prime/W\nonumber\\[.2cm]
T^3_{12}&=&B^\prime/W=k^\prime/W.
\label{torcom}
\end{eqnarray}
In this spacetime 1-forms are
\begin{eqnarray}
\theta^0&=& dx^0+A dx^2\nonumber\\
\theta^3&=& dx^3+B dx^2\nonumber\\
\theta^1&=& dx^1\nonumber\\
\theta^2&=& W dx^2
\end{eqnarray}
where $(x^0,x^1,x^2,x^3)$ corresponds to $(t,z,r,\phi)$
and the connection 1-forms are
\begin{eqnarray}
\omega^0_3&=& {f_0\over W}\theta^2={f_3\over W}\theta^2\nonumber\\
\omega^0_1&=&0\nonumber\\
\omega^0_2&=& -{f_0\over W}\theta^0\nonumber\\
\omega^3_1&=& 0\nonumber\\
\omega^3_2&=& -{h_3\over W}\theta^3\nonumber\\
\omega^1_2&=& -{W^\prime\over W}\theta^2.
\end{eqnarray}
Therefore, corresponding energy momentum components become
\begin{eqnarray}
G_{00}&=&-({h_3\over W}^2+{A\over W}{h_{03}\over W}+{W^{\prime\prime}\over W})\nonumber\\[.2cm]
G_{33}&=&{f_0\over W}^2+{B\over W}{f_{03}\over W}+{W^{\prime\prime}\over W}\nonumber\\[.2cm]
G_{11}&=&-[{f_0h_3\over W^2}+{f_0\over W}^2+{h_3\over W}^2+{f_{03}\over W}{B\over W}+{h_{03}\over W}{A\over W}]\nonumber\\[.2cm]
G_{22}&=&-{f_{0}h_3\over W^2}\nonumber\\[.2cm]
G_{01}&=&-{f_{0}q^\prime\over W^2}\nonumber\\[.2cm]
G_{02}&=&{f_{33}\over W}\nonumber\\[.2cm]
G_{31}&=&-{h_{3}k^\prime\over W^2}\nonumber\\[.2cm]
G_{32}&=&{h_{00}\over W}\nonumber\\[.2cm]
G_{03}&=&(f_0-h_3){h_0\over W^2}\nonumber\\[.2cm]
G_{12}&=&-(f_0+h_3){W^\prime\over W^2}
\label{enmom3}
\end{eqnarray}
with the condition $f_3=h_0$. $f_0=\partial f/\partial x^0$, etc., and a prime denotes a partial  derivative with respect to $x^1$.
If we take $h_3=f_0=0$ in (\ref{enmom3}), i.e., $f$ and $h$ are
functions of $z$ and $t$ respectively, we conclude that only
non-zero components of the $G_{ij}$   are
\begin{eqnarray}
G_{00}=G_{33}=-{W^\prime\over W}
\end{eqnarray}
which represents a spinning cosmic string with radius $r_0$ and its tension
is equal to its linear rest energy.

For the general form of the $f$ and $h$ space-time represents an
anisotropic spin fluid with the velocity ${\bf u}=(u^0,u^3,0,0)$,
pressures $p$ (in the $r$ and $\phi$ directions) and $P$ (in the $z$
direction) and energy momentum components are defined as
\begin{equation}
t_{ij}=(\epsilon+p)\,u_i u_j-p\,\eta_{ij}+(P-p)\,v_iv_j+2u_j\, u^k\, {\dot S}_{ki}
\label{enflu}
\end{equation}
where $\epsilon$ is the rest energy of the spin fluid, ${\bf v}$ is
a spacelike vector orthogonal to timelike velocity vector $u$ and
antisymmetric part of the hypermomentum density is equal to the spin
angular momentum $S_{ij}$ \cite{ob}. Since only $S_{12}$, $u^0$ and
$u^3$ are different from zero the Frenkel conditions automatically
are satisfied and the last term  in  (\ref{enflu}) does not
contribute to the energy momentum tensor. Combined energy-momentum
equations are given as
\begin{eqnarray}
G_{00}&=&-({h_3\over W}^2+{A\over W}{h_{03}\over W}+{W^{\prime\prime}\over W})
=(\epsilon+p)u_0u_0+p+(P-p)v_0v_0\nonumber\\[.2cm]
G_{33}&=&{f_0\over W}^2+{B\over W}{f_{03}\over W}+{W^{\prime\prime}\over W}
=(\epsilon+p)u_3u_3-p+(P-p)v_3v_3\nonumber\\[.2cm]
G_{11}&=&-[{f_0h_3\over W^2}+{f_0\over W}^2+{h_3\over W}^2+{f_{03}\over W}{B\over W}+{h_{03}\over W}{A\over W}]=-p+(P-p)v_1v_1\nonumber\\[.2cm]
G_{22}&=&-{f_{0}h_3\over W^2}=-p+(P-p)v_2v_2\nonumber\\[.2cm]
G_{01}&=&-{f_{0}q^\prime\over W^2}=(P-p)v_0v_1\nonumber\\[.2cm]
G_{02}&=&{f_{33}\over W}=(P-p)v_0v_2\nonumber\\[.2cm]
G_{31}&=&-{h_{3}k^\prime\over W^2}=(P-p)v_3v_1\nonumber\\[.2cm]
G_{32}&=&{h_{00}\over W}=(P-p)v_3v_2\nonumber\\[.2cm]
G_{03}&=&(f_0-h_3){h_0\over W^2}=(P-p)v_0v_3\nonumber\\[.2cm]
G_{12}&=&-(f_0+h_3){W^\prime\over W^2}=(P-p)v_2v_1
\end{eqnarray}
We have also  normalization $-u_0^2+u_3^2=-1$,
$-v_0^2+v_1^2+v_2^2+v_3^2=1$ and  orthogonality conditions
$u_0v_0=u_3 v_3$. It is seen that since $u^0$ and $u^3$ are
different from zero, fluid's velocities $u_t$, $u_z$ and $u_\phi$
are different from zero and there is no movement in $r$ direction as
expected.

In the space-time with torsion junction conditions are different
from \ref{subcon} and here they obey the the matching conditions
given by \cite{akp}
\begin{eqnarray}
g_{\mu\nu}\vert_-&=&g_{\mu\nu}\vert_-\nonumber\\[.4cm]
{\partial g_{\mu\nu}\over\partial x^\alpha}\vert_-+2 g_{\alpha\rho} K^\rho_{(\mu\nu)}\vert_-&=&
{\partial g_{\mu\nu}\over\partial x^\alpha}\vert_-+2 g_{\alpha\rho} K^\rho_{(\mu\nu)}\vert_-
\end{eqnarray}
on the cylinder with radius $r=r_0$ and constant $t=t_0$, $z=z_0$.
For our example, these  conditions reduce to
\begin{eqnarray}
g_{\phi\phi}\vert_+&=&g_{\phi\phi}\vert_-\nonumber\\[.2cm]
g_{\phi z}\vert_+&=&g_{\phi z}\vert_-\nonumber\\[.2cm]
g_{\phi t}\vert_+&=&g_{\phi t}\vert_-\nonumber\\[.2cm]
g_{\phi\phi,r}\vert_+&=&g_{\phi\phi,r}\vert_-+2K_{\phi\phi}^r\nonumber\\[.2cm]
g_{\phi z,r}\vert_+&=&g_{\phi z,r}\vert_-+2K_{(\phi z)}^r\nonumber\\[.2cm]
g_{\phi t,r}\vert_+&=&g_{\phi t,r}\vert_-+2K_{(\phi t)}^r\nonumber\\[.2cm]
\label{jun}
\end{eqnarray}
where $(_+)$ represents torsionless exterior space-time and $(_-)$
represents the interior space-time  with ${K_{\mu\nu}}^\rho$ which
is the contorsion part of the connection
${\Gamma_{ij}}^k={\Gamma_{ij}}^{\{\}k}-{K_{ij}}^k$. Symmetric part
of the connection ${\Gamma_{ij}}^{\{\}k}$ represents the Riemannian
contribution and the other quantities can be defined as
\begin{eqnarray}
g_{ij}&=&e_i^\alpha e_j^\beta \eta_{\alpha\beta}\nonumber\\[.2cm]
K_{kij}&=&e_i^\alpha e_j^\beta K_{k\alpha\beta}\nonumber\\[.2cm]
T^i&=&K^i_j\wedge\theta^j\nonumber\\[.2cm]
K^i_j&=&{{K_k}^{\alpha}}_\beta dx^k
\end{eqnarray}
And if we rewrite conditions (\ref{jun}) explicitly
\begin{eqnarray}
-a^2 z_0^2+a^2 t_0^2+b^2 r_0^2&=&-A^2+B^2+W^2\\[.2cm]
a z_0&=&A\\[.2cm]
a t_0&=&B\\[.2cm]
2b^2 r_0-(-AA_r+BB_r+WW^\prime)&=&AA_r-BB_r\label{ek}\\[.2cm]
A_r&=&A_r\label{ek1}\\[.2cm]
B_r&=&B_r
\label{jun1}
\end{eqnarray}

Some exact solutions of Einstein field equations have CTCs. It means
that a timelike observer instead of going from past to future
intersects itself. In \cite{bonnor} Bonnor suggets that instead of
spacetime, a closed timelike curve may exist in the laboratory by
considering two spinning massive particles having angular momenta
and parallel spins. It is shown that a CTC can be created in the lab
depending on the particles' unit angular momenta.

The spacetime given with (\ref{genspin}) has CTCs in the region
\begin{eqnarray}
r&>&\sqrt{F^2-B^2}/b\nonumber\\[.2cm]
&=&\left[a(z+t)(J^t+J^z)\right]\left[a(z-t)(J^t-J^z)\right]
\end{eqnarray}
which changes the sign of $g_{\phi\phi}$ for $F^2>B^2$ and spacetime
has global hyperbolic structure for $F^2<B^2$.

If we consider $z_0=const.$ hypersurface, spin and dislocation
parameters  $J^t$ and $J^z$  play role in formation of the CTCs.

\section{Conclusion}
In this work, a general form of the spinning cosmic string in the
Einstein theory of gravity is presented and the same problem is
considered in the Einstein Cartan theory of gravity. Our metric is
studied in the EC theory and as pointed out in \cite{punsol} and
\cite{tod} besides the curvature and energy-momentum singularity
space-time has the torsion singularity characteristic. The exact
solution of this space-time is considered as a fluid which is
non-zero source in Einstein theory and  spin fluid with anisotropic
pressure moving along the symmetry axis in EC theory.

Since it is important to understand CTC's role in the general
relativity, CTCs are calculated in the spacetime suggested. Our
solution can also be expanded multi-string solutions by redefining
coordinate parameters\cite{gallet,let}.
\\[2mm]
\noindent {\bf {Acknowledgements}} \\ [2mm] \noindent Author would
like to thank W.F. Hehl, B. Mashhoon, A.N. Aliev and M. Horta\c csu
for valuable discussions and suggestions. This work partially
supported by TUBITAK.
\\[4mm]

\end{document}